\documentclass[prl,twocolumn,showpacs,superscriptaddress,floatfix]{revtex4}

\usepackage{amsmath}
\usepackage{graphicx}
\usepackage{dcolumn}
\usepackage{bm}
\begin{document}
{submitted to PRL'11}

\title{Polariton ring condensates and sunflower ripples in an expanding quantum liquid}
\author{Gabriel Christmann}
\email{gprmc2@cam.ac.uk}
\affiliation{NanoPhotonics Centre, Cavendish Laboratory, University of Cambridge, J.\ J.\ Thomson Avenue, Cambridge CB3 0HE, England, UK.}%
\author{Guilherme Tosi}
\affiliation{NanoPhotonics Centre, Cavendish Laboratory, University of Cambridge, J.\ J.\ Thomson Avenue, Cambridge CB3 0HE, England, UK.}%
\altaffiliation{Departamento de Fisica de Materiales, Universidad Autonóma, E28049 Madrid, Spain.}

\author{Natalia G. Berloff}
\affiliation{Department of Applied Mathematics and Theoretical Physics, University of Cambridge, Cambridge, CB3 0WA, UK.}%

\author{Panos Tsotsis}
\affiliation{Department of Materials Science and Technology, University of Crete, P.O. Box 2208, 71003 Heraklion, Greece.}

\author{Peter S. Eldridge}
\author{Zacharias Hatzopoulos}
\affiliation{FORTH-IESL, PO Box 1527, 71110 Heraklion, Crete, Greece.}

\author{Pavlos G. Savvidis}
\affiliation{Department of Materials Science and Technology, University of Crete, P.O. Box 2208, 71003 Heraklion, Greece.}
\affiliation{FORTH-IESL, PO Box 1527, 71110 Heraklion, Crete, Greece.}

\author{Jeremy J. Baumberg}%
\affiliation{NanoPhotonics Centre, Cavendish Laboratory, University of Cambridge, J.\ J.\ Thomson Avenue, Cambridge CB3 0HE, England, UK.}%

\date{\today}

\begin{abstract}
Optically pumping high quality semiconductor microcavities allows for the spontaneous formation of polariton condensates, which can propagate over distances of many microns. 
Tightly focussed pump spots here are found to produce expanding incoherent bottleneck polaritons which coherently amplify the ballistic polaritons and lead to the formation of unusual ring condensates.
This quantum liquid is found to form a remarkable sunflower-like spatial ripple pattern which arises from self interference with Rayleigh-scattered coherent polariton waves in the \v{C}erenkov regime.
\end{abstract}

\maketitle


Condensing solid state bosonic quasiparticles brings prospects for macroscopic quantum-coherent integrated devices. Although still cryogenic, advances in coherent atom matter-wave optics \cite{Anderson98,Cataliotti01} shows what might be possible with room temperature condensation, as sought for years with superconductors. In the last decade it has become apparent that polaritons formed from coupling semiconductor excitons with cavity photons produce bosonic quasiparticles that can condense \cite{Deng02,Kasprzak06}, even at room temperature \cite{Christopoulos07}, with superfluid transport and a number of unusual vortex states now discussed \cite{Lagoudakis08,Amo09,Amo09b}.

Polariton condensation requires high-$Q$ optical cavities, large light-matter Rabi coupling exceeding thermal energies ($\hbar \Omega \! > \! k_B T$), and efficient relaxation into bosonic quasiparticles from the high-energy injected fermionic carriers. Successful condensation in microcavities made of III-V arsenides \cite{Deng02}, II-VI tellurides \cite{Kasprzak06}, III-V nitrides \cite{Christopoulos07,Christmann08} and anthracene \cite{Kena10} have relied on natural spatial traps produced by the disordered energy landscape to confine polaritons sufficiently to initiate condensation, as well as artificially created ones. This is largely due to the existence of a `bottleneck' \cite{Kavokin07} in the relaxation pathway where polaritons collect, thus forming a high-energy reservoir from which they cannot relax further. Condensation requires an additional localized concentration of the bosons. However such disorder-induced localization severely hampers development of chip-based condensate circuits. 


Here we demonstrate an efficient scheme to pump such non-equilibrium quantum liquids, that produces unusual coherent spatial ring states. These macroscopic condensate rings experience spatially-localized stimulated scattering as the bosons expand ballistically. Despite their outwards acceleration the quantum liquid remains phase coherent. We show that spatial ripples appearing as interferometric sunflowers arise from elastic scattering in the superfluid.

\begin{figure*}[tb]
\includegraphics[width=17cm]{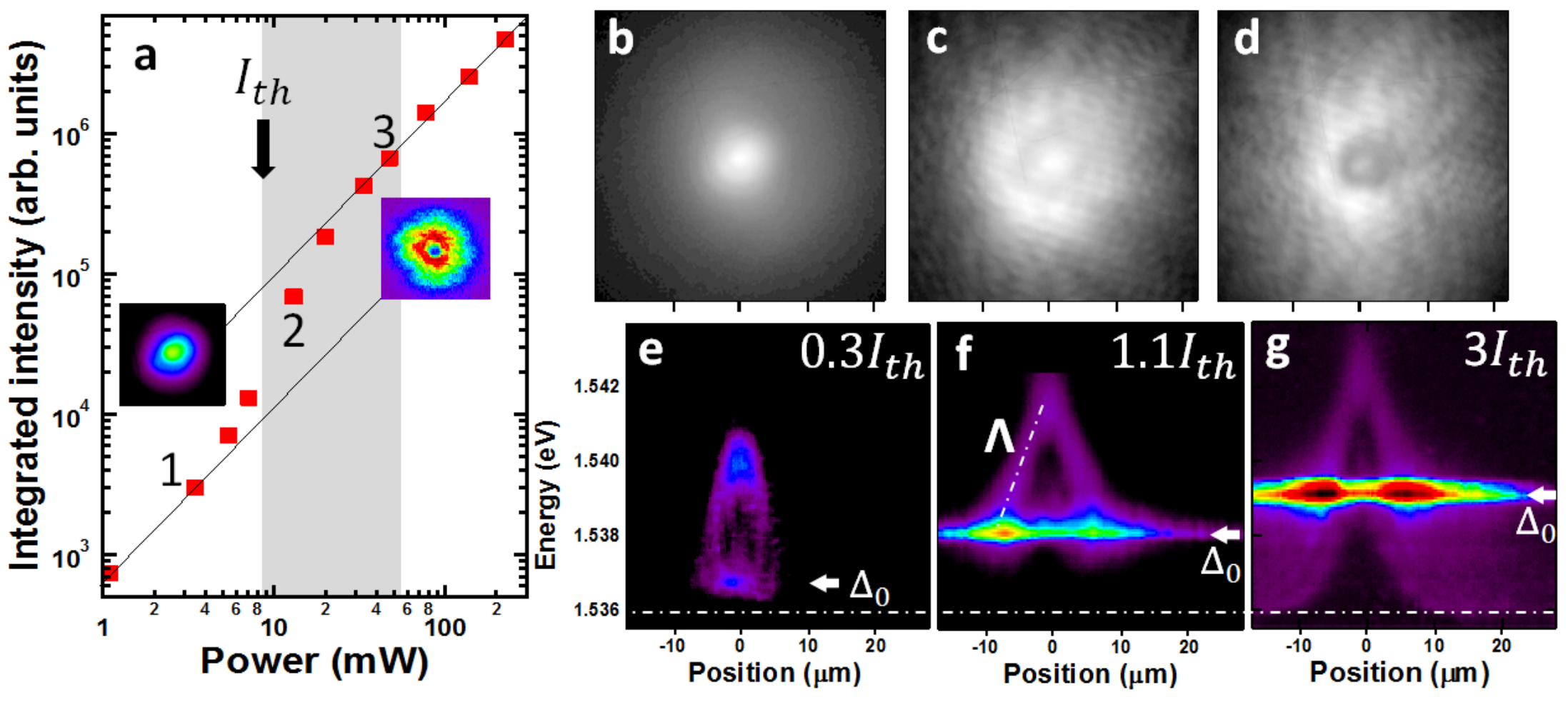}
	\caption{(a) Integrated polariton emission \textit{vs} pump power, insets are real space emission images. Black lines are guide to the eye above/below threshold at 9mW. Grey region corresponds to polariton condensate. (b-d) Real space images (log greyscale) of emission for the powers corresponding to points marked 1-3 in (a) at $I_p$=0.3, 1.1, and 3.0 $I_{th}$. (e-g) Corresponding emission intensities for 1-3 on energy \textit{vs} position maps (log scale). Horizontal dashed line is the non-blueshifted $k$=0 energy.}
\end{figure*}

The microcavities consist of a $5\lambda/2$ AlGaAs cavity containing four sets of three GaAs quantum wells placed at the antinodes of the cavity electric field, surrounded by two AlGaAs/AlAs Bragg mirrors of 35 (bottom) and 32 (top) pairs. The cavity quality factor is measured to exceed Q$>$16000, with 
cavity photon lifetime $\tau_c$=9ps. Strong coupling is obtained with a Rabi splitting between lower and upper polariton energies of 9 meV. All data presented here use a negative cavity detuning of -5meV, although other negative detunings give similar results. Excitation is provided by a single-mode narrow-linewidth CW laser focused to a 1$\mu$m diameter spot
and tuned to the first spectral dip at energies above the high-reflectivity mirror stopband at 750nm. To prevent unwanted sample heating the pump laser is chopped at 100Hz with an on:off ratio of 1:30. Both real and \textbf{k}-space images of the photoluminescence (filtering out the pump laser) can be imaged, either on the entrance slits of a spectrometer/CCD combination, or directly on a CCD.

\begin{figure}[tb]
\includegraphics[width=9cm]{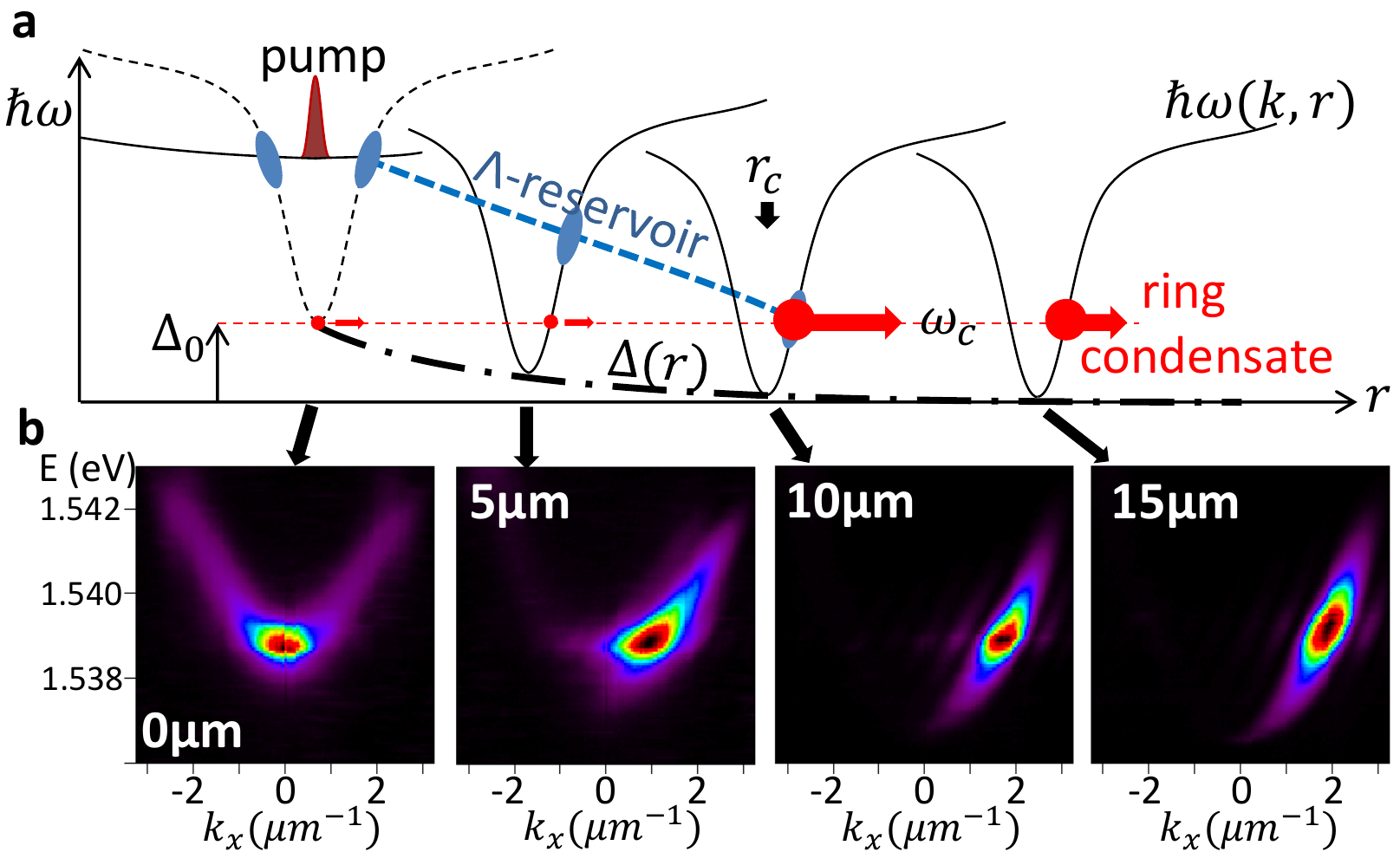}
	\caption{(a) Schematic showing dispersion relations at increasing radial position, together with (b) corresponding measured emission from dispersion relations at each position.}
\end{figure} 

We observe that the micron-scale polariton reservoir resulting from the tight pump spot efficiently feeds a macroscopic quantum state in the form of a spatial ring which appears above the threshold pump power for condensation (Fig.1a). While spatial images below threshold show maximum incoherent emission from the central pumped spot (Fig.1b), above threshold the coherent condensate emission develops in a ring surrounding the pump 10$\mu$m across (Fig.1c) independent of precise position on the semiconductor wafer,  which becomes more pronounced at high powers (Fig.1d). The entire condensate is spatially coherent (as observed also with interferometry \cite{Kasprzak06}, not shown). The smooth emission profile observed below threshold acquires a peculiar `sunflower' pattern of interference ripples above threshold, which are always observed (Fig.1c,d) and stable for many minutes.

The origin of these ring states lies in the spatially-dependent energy shifts $\Delta(r)$, which create an optically-controlled potential landscape. Polaritons are highly nonlinear quasiparticles, and experience strong mutual repulsion resulting in energy blue shifts with increasing density \cite{Kasprzak06,Tosi11}. This effect can be clearly seen in the blue shifts of the condensate energy $\omega_c = \Delta(0) = \Delta_0$ with increasing laser power (Fig.1e-g) which show the spectra emitted at different distances from the pump spot. 
Near the pump spots, above threshold emission is suppressed around the condensate energy and only at larger distances do polariton states below the condensate become available. 

%
The condensate is created at in-plane $k$=0 at the bottom of the blue-shifted polariton dispersion in the middle of the pump spot, and gains momentum as it accelerates outwards (Fig.2a). This is confirmed by resolving the emission from the dispersion relation $E(k)$ at each position (Fig.2b). These observations form the 2D counterpart to that observed in 1D polariton wires\cite{Wertz10}, leading to the formation of condensates with circular symmetry extended over $>100\mu$m$^2$.

The ring condensation mechanism is revealed by the weak uncondensed hot background polariton luminescence in the same spectra (Figs.1e-g) above the condensate energy. The $\Lambda$-shaped emission observed originates 5meV above the $k$=0 polariton and drops in energy with distance from the centre (dashed) until it meets the ring condensate exactly at its intensity maximum at $r=r_c$. 
Suprisingly, $\Lambda$-shaped emission is also observed below threshold. 
These hot polaritons possess a large momentum and come from the high momentum polariton reservoir close to the bottleneck.
Despite ballistically expanding from the pump spot, they decelarate, thus dropping down the polariton dispersion (see sequence in Fig.2a). Whilst condensed polaritons cannot inelastically scatter, the incoherent polaritons decelerate in a systematic way. Eventually this incoherent reservoir meets the energy of the condensate and amplifies it strongly, leading to a ring shaped condensate.

The ring condensate cannot be described as independently accelerating polaritons in each radial direction because of the varying $k_c(r)=K \! [\omega_c - \Delta(r)]$ set by the position dependent blueshifts.
Instead  the condensate wavefunction is a superposition of different $k$-states at different radial positions,
\begin{equation}
\psi(r,t) =  e^{ i \left\{ K \! [\omega_c - \Delta(r)].r -  \omega_c t \right\} } \, g(r) e^{-t/\tau_c(r)}
\label{eqn:cond}
\end{equation}
where $\tau_c$ is the polariton decay time, and the dispersion relation $K(\omega) = k \approx \sqrt{2 m^* \omega / \hbar}$ for small $k$. 
As the condensate expands, its velocity changes so that the distance $\rho$ is reached after a time $\tau$,
\begin{equation}
\tau = \int_0^\rho \left. \frac{d K}{d \omega} \right|_{\omega_c} dr
\label{eqn:vel}
\end{equation}

\begin{figure}[tb]
\includegraphics[width=8cm]{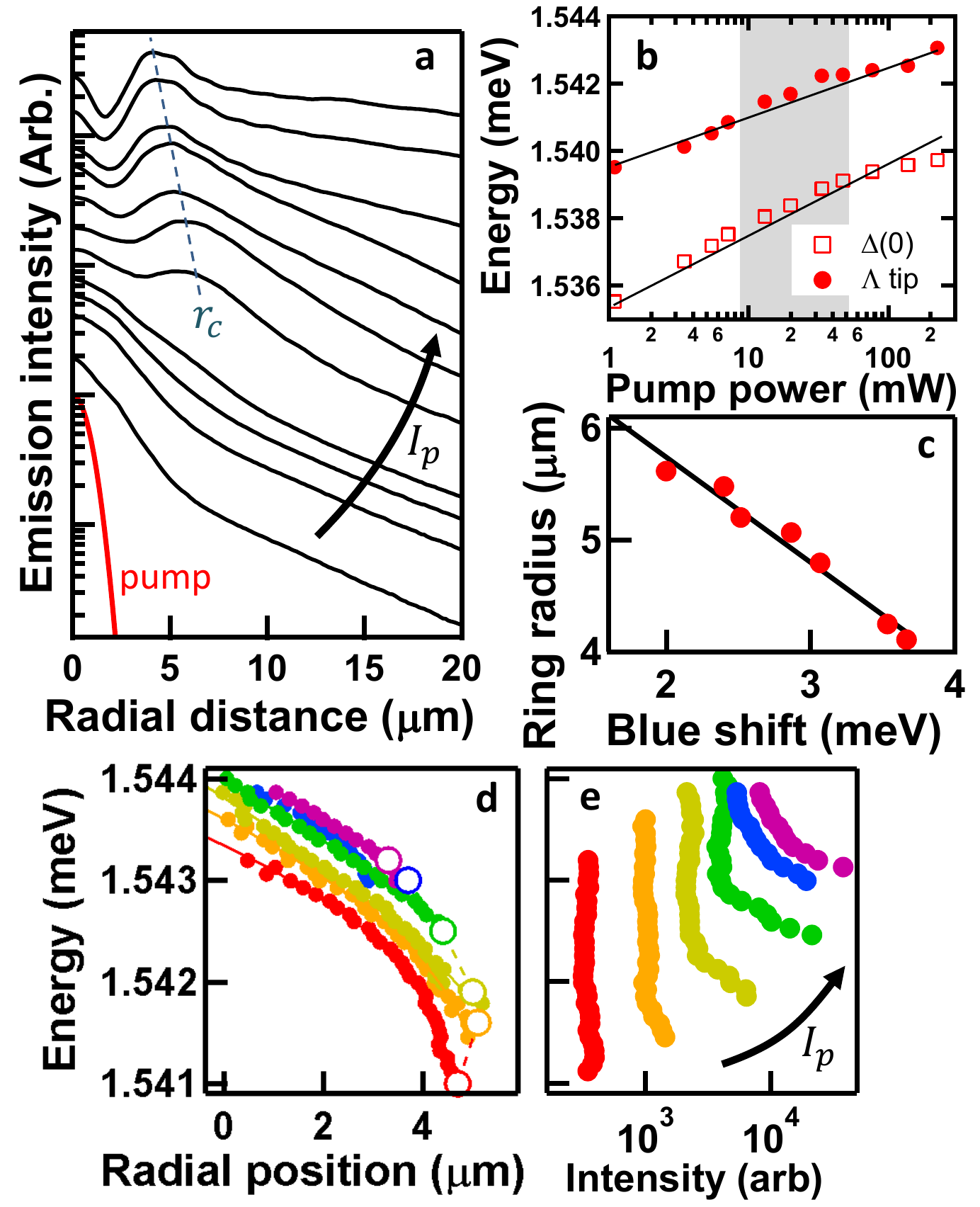}
\caption{(a) Averaged radial intensity for increasing excitation powers (black lines), with excitation profile (red line). (b) Energy of the bottom of the lower polariton branch at the excitation spot (circles) and energy of the $\Lambda$ tip (squares) \textit{vs} excitation power. (c) Distance of the ring condensate from excitation spot \textit{vs} power. Lines are guide to the eye. (d,e) Energy position and intensity of the incoherent reservoir polaritons \textit{vs} distance for increasing powers. White circles mark condensate energies (different dataset).}
\end{figure} \noindent

The amplitude $g(r)$ is strongly amplified at the critical radius $r_c$ (Fig.3a) when stimulated scattering converts incoherent polaritons from the reservoir into the coherent expanding condensate of Eq.(\ref{eqn:cond}). This unusual condensate thus has a wavevector that varies with radial position, and is a macroscopic object so that individual local scattering events are suppressed. Below threshold, incoherent polaritons building up at $k$=$r$=0 are accelerated outwards with a decay length of 2.1$\mu$m. However they experience no amplification at $r_c$ since there is no large population in any one state (they are not Bose condensed) and stimulated scattering is thus slow. At threshold, the ring immediately forms at $r_c$=5.6$\mu$m (Fig.3c) and the intensity there increases by almost ten-fold (Fig.3a). The subsequent ballistic propagation of the condensate is increased (to 4.4$\mu$m) due to the blue-shifts which move $\omega_c$ further up the dispersion giving larger outward velocities. As the excitation power increases the ring shrinks in diameter linearly with the blue shift (Fig.3c), and the ring intensity increases linearly (as Fig.1a).



This shrinking ring tracks the position at which the $\Lambda$ reservoir intersects with the increasingly blue-shifted condensate emerging from the pumped spot (Fig.3d). These extracted energies of the incoherent reservoir polaritons at increasing radial distance ($\Lambda$-shaped in Fig.1e-g) change little in shape with power. They are rigidly blueshifted, but less than the condensate mode itself (Fig.3b). Hence the reservoir polaritons are not injected at a fixed energy above the polariton dispersion, but experience thermally-induced energy changes at the pump spot (believed to be due to local heating at the very centre of the pump spot which red-shifts the exciton locally). 
A second feature is the constant deceleration rate of the expanding reservoir polariton cloud (note the lack of polariton emission from {\it inside} the $\Lambda$ line in Figs.1e-g implies that this expansion is not diffusive but ballistic). Since $\hbar \dot{k} = -F$ this implies a constant opposing force $F$ on the polaritons (as in Bloch oscillators \cite{Gluck02}). This force seems to arise from interactions of the incoherent reservoir polaritons with excitons near the pump spot. Such interactions cannot occur with the coherent polaritons because the condensate, being a single wavefunction, cannot decelerate simultaneously all around the ring through single local scattering events (as these cannot conserve energy and momentum). The preservation of incoherent polaritons throughout this decelaration process (Fig.3e) supports this scenario.

%
%
%

The characteristic sunflower self-interference ripples (Fig.4a) clearly demonstrate that a component of the quantum liquid propagates in non-radial (spiral) directions. The patterns are persistent and regular, and can be decomposed by local 2D Fourier transforms within different regions of the image. This analysis reveals that at every point the $\Delta k$ corresponding to the fringe periods and directions (formed by interference between the outwards-moving condensate and accompanying scattered waves) lies on a circle touching the origin and of radius $k_c = K(\omega_c)$  (Fig.4b). This is exactly the pattern expected for isotropic Rayleigh scattering of the outwards-travelling condensate and is experimentally observed (Fig.4c) \cite{Houdre00}.

\begin{figure}[tb]
\includegraphics[width=7cm]{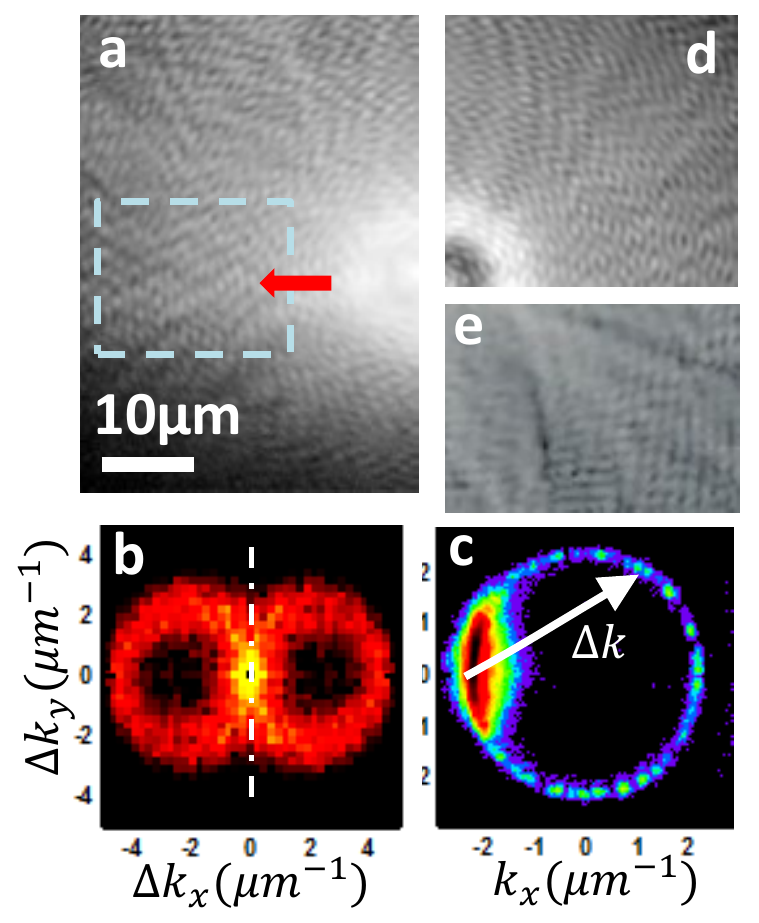}
	\caption{(a) Real space image of the emission (log scale), together with (b) Fourier transform and (c) $k$-space image spatial filtered within the rectangle marked.  (d) Simulated interference of the condensate with Rayleigh scattered waves.  (e) Simulation of the spatial polariton density using the cGL equation (see text).}
\end{figure}

Confirmation of this analysis is provided firstly by simulations that interfere a circular wave (the condensate) with weaker (Rayleigh-scattered) plane waves in random directions and with random phases (Fig.4d). The resulting image from this simple model reproduces impressively well the experimental observations, with 6\% of the total energy in the Rayleigh component. 
A more accurate theoretical description of the system starts from the mean field equation \cite{Wouters07,Wouters10,Keeling08} in the form of the cubic Ginsburg-Landau (cGL) equation, 
\begin{multline}
  \label{eq:1}
   i \hbar \frac{\partial\psi}{\partial t} = 
    \bigl[
    -
    \frac{\hbar^2 \nabla^2}{2 m} 
    + U_0|\psi|^2 + V_{\rm ext}({\bf r})
    +i P({\bf r}, \psi)  -i\kappa \bigr] \psi
\end{multline}
where $U_0$ is the strength of the interaction, and the details of pumping are included in 
$P({\bf r},\psi) = \gamma - \eta i \hbar \partial_t -\Gamma
|\psi|^2 $ (inside the pump spot only).  We take the polariton decay rate
$\kappa=0.1$meV, radius of pump spot $R$=10$\mu$m, and consider pumping at $3$ times threshold so $\gamma=3\kappa$, with $\Gamma/U_0=0.3$ where $\Gamma$ is the rate of nonlinear dissipation leading to gain saturation
 \cite{Keeling08}.  The dimensionless parameter $\eta$=0.1  describes the energy relaxation due to interactions with the reservoir particles. To demonstrate the Rayleigh scattering we assume that the external potential $V_{\rm ext}$ contains disorder and is modelled by a random distribution of Fourier modes of maximum amplitude $0.1$meV with wavelengths between $2-50\mu$m. 
The healing length in our model is given by $\ell_0 = \sqrt{\hbar^2/2 m \kappa}= 1.95 \mu$m
for $m = 10^{-4} m_e$. The random potential therefore creates obstacles of size on the order of the healing length for the polariton flow. It is well-known \cite{El06} that linear sound waves are emitted as a result of  supersonic flow around such small impurities whereas for larger obstacles the emitted waves are not linear but are dispersive shocks consisting of oblique solitons. Such Cherenkov emission of sound waves is clearly seen in Fig.4e. Observation of this pattern implies the breakdown of superfluidity in the polariton condensate. A simple estimate shows that away from the pump spot the flow velocity, ${\bf u}=\hbar \nabla \phi/m$ where $\phi$ is the phase of $\psi$, becomes much greater than the local sound velocity $c_s=\sqrt{U_0 \rho/m}$ where the density $\rho=|\psi|^2$. The Mach number, $M$, which is the ratio of  the flow velocity to the local speed of sound, is of the order of $\sqrt{\mu/U_0\rho}$, where $\mu$ is the chemical potential of the system. Neglecting the quantum pressure term it follows from (\ref{eq:1}) that $\mu=U_0(\gamma-\kappa)/(\eta U_0 + \Gamma)$, $\rho(0)=\mu/U_0$ and $\rho< \rho(0) \exp[-2\kappa (r-R)/\hbar\sqrt{\mu/m}]/(r-R)$ away from the pumping spot. It follows that $M\gg 1$  as soon as $\exp\left\{2\kappa (r-R)/\hbar\sqrt{\mu/m}\right\}(r-R)\gg 1$ which takes place within a couple of healing lengths away from the boundary of the pump spot. Comparing to experiments of Amo and cowokers \cite{Amo09b} where the superfluid is at subsonic flow velocities, here the flow velocity is supersonic which places the system in the \v{C}erenkov regime where scattering is allowed.

In conclusion, expanding ring condensates are thus clearly non-equilibrium quantum liquids, which can be efficiently excited and locally amplified using incoherent reservoirs. However they also suffer Rayleigh scattering, which leads to condensate fragmentation once the random potential exceeds a certain amplitude (thus demanding high quality microcavity samples). Our methods finally allow the generation of 2D condensates in arbitary locations, which can be spatially manipulated for use in coherent solid state devices, and open up the possibility for electrically-pumped condensate chips.


We acknowledge EPSRC grants EP/G060649/1, EP/F011393, EU CLERMONT4 PITNGA-2009235114 and INDEX PITNGA-2011-289968 and Spanish MEC (MAT2008-01555). G.T. acknowledges FPI scholarship support from the Spanish MICINN.


\begin{thebibliography}{bib}

\bibitem{Anderson98} B.P. Anderson, M.A. Kasevich, Science {\bf 282}, 1686 (1998).
\bibitem{Cataliotti01}  F.S. Cataliotti {\it et al.}, Science {\bf 293}, 843 (2001).


\bibitem{Deng02} H. Deng {\it et al.} Science {\bf 298}, 199-202 (2002).


\bibitem{Kasprzak06} J. Kasprzak \textit{et al.} Nature \textbf{443} 409-414 (2006).

\bibitem{Tosi11} G. Tosi \textit{et al.}, Nat. Phys. (2011).


\bibitem{Christopoulos07} S. Christopoulos \textit{et al.} Phys. Rev. Lett. \textbf{98}, 126405 (2007).


\bibitem{Lagoudakis08} K. G. Lagoudakis \textit{et al.}  Nat. Phys. \textbf{4}, 706-710 (2008).


\bibitem{Amo09} A. Amo \textit{et al.}, Nature \textbf{457}, 291 (2009).


\bibitem{Amo09b} A. Amo \textit{et al.}  Nat. Phys. \textbf{5}, 805 (2009).





\bibitem{Christmann08} G. Christmann \textit{et al.} Appl. Phys. Lett. \textbf{93}, 051102 (2008).

\bibitem{Kena10} S. K\'ena-Cohen, S.R. Forrest Nat. Photonics \textbf{4}, 371-375 (2010).

\bibitem{Kavokin07} A.V. Kavokin, J.J. Baumberg, G. Malpuech, F.P. Laussy, {\it Microcavities} (OUP, Oxford, 2007).


\bibitem{Wertz10} E. Wertz \textit{et al.} Nat. Phys. \textbf{6}, 860 (2010).






\bibitem{Amo10} A. Amo, \textit{et al.} Phys. Rev. B \textbf{82}, 081301 (2010).

\bibitem{Gluck02} M. Glück, A.R. Kolovsky, H.J. Korsch, Phys. Rep. {\bf 366}, 103 (2002).


\bibitem{Houdre00} R. Houdr\'e \textit{et al.} Phys. Rev. B \textbf{61}, R13333 (2000).


\bibitem{Wouters07} M. Wouters, and I. Carusotto Phys. Rev. Lett. \textbf{99}, 140402 (2007).

\bibitem{Wouters10} M. Wouters and I. Carusotto, Phys. Rev. Lett. {\bf 105}, 020602 (2010).

\bibitem{Keeling08} J. Keeling, N. G. Berloff, Phys. Rev. Lett. \textbf{100}, 250401 (2008).

\bibitem{El06} G.A. El and A.M. Kamchatnov, Phys Letts A {\bf 350}, 192 (2006).





\end{thebibliography}
\end{document}